\documentclass[11pt]{article}
\usepackage[latin1]{inputenc}
\usepackage{graphicx}
\usepackage[numbers,sort&compress,round]{natbib}

\usepackage{dcolumn}
\usepackage{bm}
\usepackage[usenames]{color}

\begin{document}

{\Large \noindent Geometry-induced asymmetric diffusion}\\

{\large \noindent Robert S. Shaw$^\ast$, Norman Packard$^{\ast,\dag,\ddag}$, Matthias Schr\"oter$^\S$, Harry L. Swinney$^\S$}\\

\noindent$^\ast$ProtoLife Srl, via della Libert\'{a} 12, 30175 Venezia, Italy,
$^\dagger$European Center for Living Technology, S. Marco 2847, 30124 Venezia, Italy,
$^\ddagger$Santa Fe Institute, 1399 Hyde Park Road, Santa Fe, NM 87501,
$^\S$Center for Nonlinear Dynamics and Department of Physics,
 University of Texas at Austin, Austin, TX 78712
\vspace{2mm}

{\bf \noindent
Past work has shown that ions can pass through a membrane more
readily in one direction than the other.  We demonstrate here in a
model and an experiment that for a mixture of small and large
particles such {\em asymmetric diffusion} can arise solely from an
asymmetry in the geometry of the pores of the membrane. Our
deterministic simulation considers a two-dimensional gas of elastic
disks of two sizes diffusing through a membrane, and our laboratory
experiment examines the diffusion of glass beads of two sizes
through a metal membrane. In both experiment and simulation, the
membrane is permeable only to the smaller particles, and the
asymmetric pores lead to an asymmetry in the diffusion rates of
these particles. The presence of even a small percentage of large
particles can clog a membrane, preventing passage of the small
particles in one direction while permitting free flow of the small
particles in the other direction.  The purely geometric 
kinetic constraints may play a role in common biological 
contexts such as membrane ion channels.  }

\vspace{3mm}
{\noindent asymmetric diffusion $\vert$ channels $\vert$ asymmetric pores}
\vspace{3mm}

Asymmetric diffusion appears in studies of ion transport across
membranes
\cite{hille:01,chen:93,siwy:02,siwy:05,siwy:06,kosinska:05,kosztin:04,cervera:05,cervera:06,Bauer:06,smeets:06,kolomeisky:07},
and in the context of osmosis \cite{chou:99,kim:05}.  
Ion pores in biological
membranes often act as effective rectifiers, blocking the passage of
a particular ion in one direction while allowing free flow in the
other.  Explanations of these phenomena emphasize electrostatic
effects or conformational changes.  For example, a blocking ion
might pass part way through a membrane pore, and then become bound,
preventing passage of subsequent ions.  If the pore is asymmetric,
the membrane can act as a rectifier. Here we demonstrate that
asymmetric diffusion can occur with no ionic binding, in a fixed
pore geometry. Blocker particles can be held in place by purely
geometrical and kinetic constraints, even though particles are in
rapid random motion.

We present both a laboratory experiment and a computer simulation to
demonstrate this effect.  In the experiment, macroscopic (1 mm)
glass beads diffuse through a metal asymmetric membrane and random
motions are generated by rapidly vibrating the container.  In the
simulation, a two-dimensional gas of elastic hard disks of different sizes
diffuses through an asymmetric membrane. In both cases, nearly
complete rectification is easily demonstrated.

\vspace{3mm}
{\noindent  \large \bf Results}
\vspace{1mm}

\begin{figure}[h]
\begin{center}
\includegraphics[width=8.7cm]{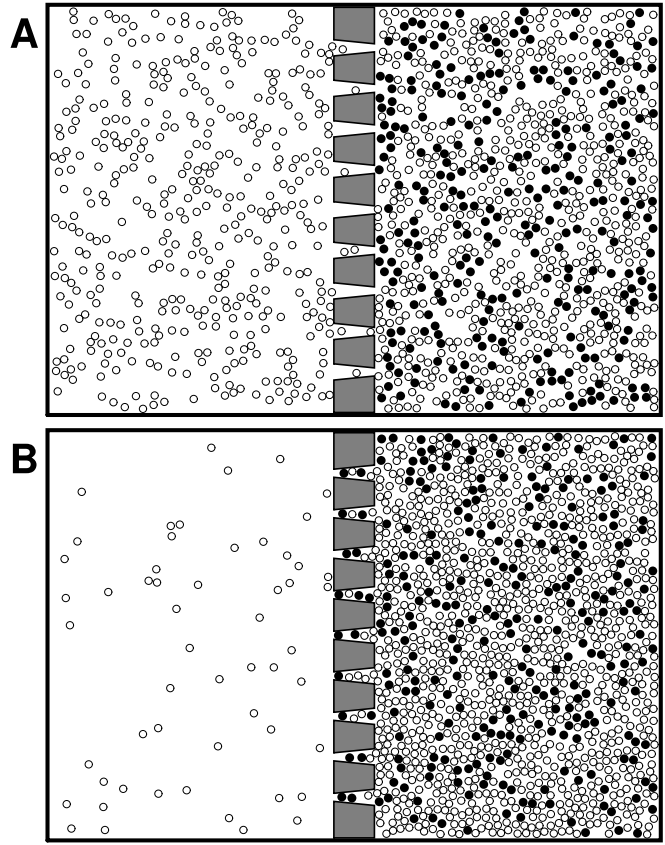}
\caption{Snapshots of a binary
mixture of hard sphere disks, initially all on the right side, that
diffuse through an asymmetric membrane whose pores are smaller on
the right end in {\it A} and on the left end in {\it B}. 
Flow through the membrane is highly reduced with the 
membrane geometry of the lower panel.
The larger particles in the gas mixture are slightly too large to pass through
the pores, and the smaller particles are just small enough to pass.
The left chamber is initially empty, and the elapsed time is the
same for {\it A} and {\it B}.} 
\label{fig:simulation_image}
\end{center}
\end{figure}

{\noindent \bf Simulations.}
Figure \ref{fig:simulation_image} shows typical snapshots from the
hard disk gas simulation. In both panels the gas is initially
restricted to the right chamber, and then tends toward equilibrium
as the smaller disks move from right to left through the membrane.
But in the lower panel, the larger disks soon clog the pores and
greatly reduce the rate of diffusion of the smaller disks through
the membrane. The presence of the smaller disks bombarding the
larger disks within the pore makes it highly unlikely that the
larger disks will be able to emerge from the pore. 

\begin{figure}[h]
\begin{center}
\includegraphics[width=8.7cm]{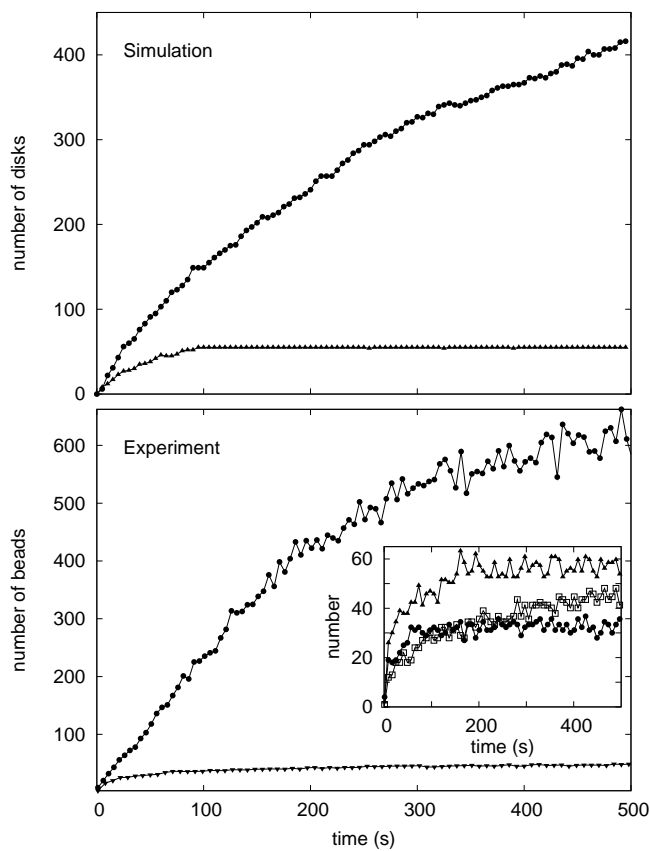}
\caption{
    The number of particles in the (initially empty) left-hand chamber (cf. Fig.~\ref{fig:simulation_image})
    in simulation and experiment.  In each graph, the upper curves correspond to pores
    with their small end on the
    right, and the lower curves correspond
    to the reverse orientation of the pores.
    The upper curves show a large diffusion rate
    (slope of the curves), while the lower
    curves illustrate that the pores rapidly clog and the diffusion rate approaches zero. The inset 
    in the lower graph shows the clogging for three different experimental
    trials.  The right-hand chamber initially had 1350 disks in
    the simulation, and 2000 beads in the experiment, which used a membrane of thickness 9.53 mm; the 
    data shown are an average of three experiments.
  In both cases 20\% of the particles
    were large and the remainder small. }
\label{fig:both_freeflow_clogging}
\end{center}
\end{figure}

Figure \ref{fig:both_freeflow_clogging} displays the concentration
in the left chamber as a function of time for both orientations of
the membrane. The flow entering from the smaller end of the pores
(Fig.~\ref{fig:simulation_image}{\it A}) behaves as if there were a
well-defined fixed diffusion constant, but the flow through the
membrane that enters the larger end of the pores
(Fig.~\ref{fig:simulation_image}{\it B}) drops to near zero, as the
membrane seals itself.  The clogging behavior of this direction of
flow is maintained even at lower concentrations of large disks. The
number of disks in the plateau region may vary from run to run,
depending on how many small disks are able to cross before the
membrane seals.

The simulated model is a purely energy-conserving Hamiltonian
system, yet it can relax to a long-lived macroscopic state of
disequilibrium, with two subsystems at different temperatures.
The more rapidly
moving disks preferentially pass through the membrane, resulting in
a higher temperature on the left side. Such a temperature difference
is typical of an effusion process\cite{cleuren:06}.
There
is a very slow equalization of temperature through collisions at the
small ends of the pores, and of concentration through the occasional
unblocking of a pore due to fluctuations, but the disequilibrium can
persist for very long times.

\clearpage

\begin{figure}[h]
  \begin{center}
  \includegraphics[width=9cm]{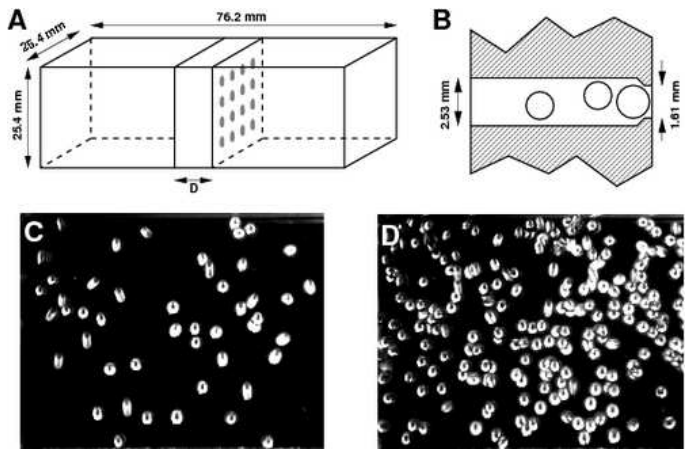}
    \caption{
    ({\it A}) Diagram of the experimental system. ({\it B}) Cross-section of a
    circular pore that has a reduced diameter on the right end, too small for the larger particles
    in the binary mixture to pass. ({\it C}) Image for 50 beads in a chamber. ({\it D}) Image for
250 beads in a chamber. }
    \label{fig:apparatus}
  \end{center}
\end{figure}

\vspace{3mm}
{\noindent \bf Experiments.}
The experiment has two species of glass beads that correspond to the two 
species of disks in the simulation, and a brass membrane with asymmetric pores, 
as illustrated in Fig.~\ref{fig:apparatus}.  The results from experiment are 
qualitatively the same as those from simulation
(Fig.~\ref{fig:both_freeflow_clogging}). Once again, there is
free flow through the membrane from the side with the small pore
ends, and clogging for flow in the other direction. The inset shows
that the number of small beads passing through the membrane before
it clogs can again vary from run to run. The clogging is not
permanent. Runs longer than 30 min show a leak rate of about 0.1
bead per minute per pore, for this number of beads.

\begin{figure}[ht]
  \begin{center}
  \includegraphics[width=8.7cm]{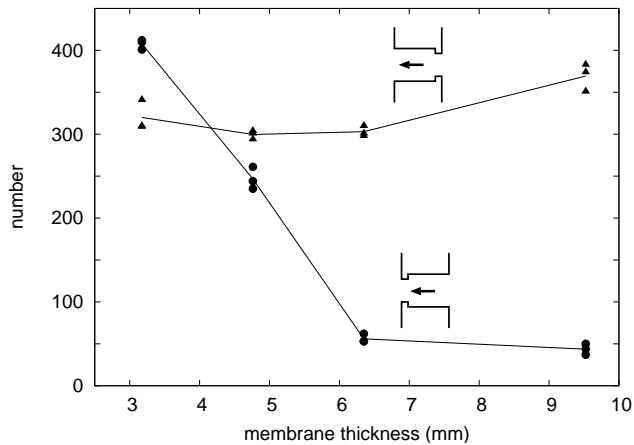}
    \caption{
    Number of small beads measured to cross the membrane after three minutes, as a function of membrane
    thickness, for the two flow directions (cf.
    Fig.~\ref{fig:apparatus}).  The loaded side of the membrane
    initially had a mixture of 1600 small beads  and
    400 large beads.}
    \label{fig:membrane_thickness}
  \end{center}
\end{figure}

Figure \ref{fig:membrane_thickness} shows the effect of membrane
thickness on the blocking effect. For the thinnest membrane, there
is no blocking effect, and the flow is even enhanced in the funneled
direction.  This is in accord with the single-species small bead
experiment to be mentioned below. The membrane must be of a minimum
thickness to produce a significant {\em backing up} of the blocking
beads. If there is a large concentration gradient across the pore,
the blocking beads will be held in place by this gradient. The
backing up requirement is central to the effect.

\begin{figure}[ht]
  \begin{center}
  \includegraphics[width=8.7cm]{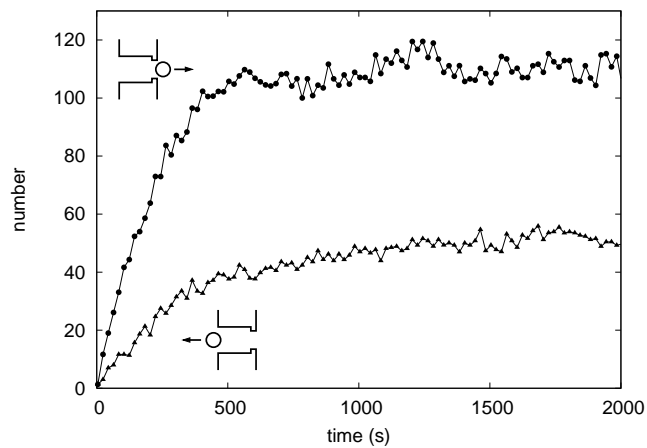}
    \caption{
    Number of beads measured to pass through a membrane (9.53 mm thick), as a function of time, when only small
    beads were present. Flow in the direction from the large to the small end of the pores is enhanced relative
    to that for the
    reverse orientation of the pores.     There was a total of 200 beads; those
    residing in the pores were not observed.}
    \label{fig:experiment_monodisperse}
  \end{center}
\end{figure}

When only small
beads are present in the experiment, there is a bias in the
dynamics, as Fig.~\ref{fig:experiment_monodisperse} illustrates.
The small beads pass more readily though the membrane
 when they enter the large end
of the pores, and the system approaches a dynamic equilibrium with
more beads on the side with the small pore ends. 
This asymmetry can be explained by the dissipative nature of the particle-wall 
collisions, which reduces the normal component of the colliding particle;
therefore the angle between the particle velocity vector and the 
sidewall becomes smaller \cite{brilliantov:04}. 
In this way the collisions inside  the large end of the pore focus a particle trajectory
towards the small end of the pore and therefore increase the effective ``cross section''.

\begin{figure}[ht]
\includegraphics[angle=-90,width=9.5cm]{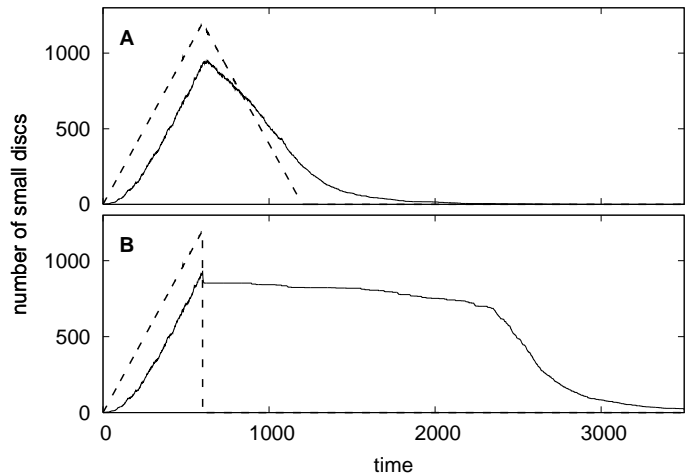}
\caption{Time-dependent variation of concentration of small disks
(solid lines) in the chamber on the right (the ``inside") in
Fig.~\ref{fig:simulation_image}{\it B}, in response to variation of
number of small disks in the left-hand (``outside") chamber (dotted
lines). The number of large disks in the inside chamber is 200 and
is zero in the outside chamber. {\it A} When the number of disks
outside is linearly increased to 1200 and then decreased linearly at
the same rate, the number of disks on the inside adiabatically
follows. {\it B} When the number of small disks outside is linearly
increased as in {\it A} and then suddenly decreased to zero, the
membrane pores clog as a consequence of the sudden strong
cross-membrane gradient, and the number of small particles on the
inside remains nearly constant for a long time.}
\label{fig:simulation_time_dependent}
\end{figure}

\vspace{3mm}
{\noindent \bf Leaky membrane simulation results.}
Some insight into the dynamics of the hard disk system can be
garnered by considering the response to an imposed variation in the
number of small particles in the chamber on the membrane side with
the small pores (left side in Fig.~\ref{fig:simulation_image}{\it
B}). The resultant macroscopic state of the chamber on the right
side, which we will call the ``inside", shows a path dependence, as
illustrated in Fig.~\ref{fig:simulation_time_dependent}. Initially
the system had 200 large particles in the inside (right) chamber and
no large particles on the outside (left) chamber; initially no small
particles were in either chamber. In
Fig.~\ref{fig:simulation_time_dependent}{\it A}, the number of small
disks in the outside chamber is gradually increased at a rate of two
disks per time unit, up to 1200 disks. Each disk is added with unit
speed, at a random position and moving in a random direction.  The
number of small disks on the inside chamber can be seen to follow
this increase, as one would expect from a simple diffusion process.
Then the number of small disks on the outside is reduced at the same
slow rate, and again the disks on the inside roughly follow. When
there is no strong concentration gradient across the membrane, the
pores conduct freely in both directions.

In Fig.~\ref{fig:simulation_time_dependent}{\it B}, the number of
small disks in the outside chamber is again increased at a rate of
two disks per time unit up to 1200 disks, and then this time the
number of disks on that side is suddenly reduced to zero; any small
disks leaking through from the inside to outside are removed from
the system. In this case, the instantaneous formation of a large
concentration gradient across the membrane causes it to clog, and
diffusion from the inside initially drops to nearly zero, even
though the final concentration outside is the same as in
Fig.~\ref{fig:simulation_time_dependent}{\it A}.

The diffusion outward does not, however, remain at zero, because the
initial concentration of blocker disks inside is small enough so
that there is a significant leakage current.  Each small step in the
descending portion of the curve in the lower panel corresponds to a
single pore opening for a time; a large blocker disk is dislodged by
a fluctuation, and a ``patch clamp" observation of that pore would
show a current of small disks flowing outward until the pore is
resealed by another large disk entering it. Eventually, the
concentration of small disks on the right becomes too small to
maintain the clogged membrane, and the concentration curve on the
right returns to a simple diffusive decrease.  Because there are
only nine pores in this simulation, the length of the plateau
corresponding to the clogging is variable, but the general shape of
the knee is repeatable.

The possible biological relevance of this clogging effect to
membrane rectification is evident. Possibly the simple rectification
described here could have played a role in tide-pool scenarios of
the origin of life.  In our simulation, repeated sawtooth driving on
the outside will maintain a concentration gradient indefinitely,
even with a leaky membrane. Similarly, one can imagine that repeated
evaporation and flooding of a primeval tide-pool may have provided
the conditions for a sustained gradient across the membrane of an
early vesicle, enough to drive a proto-metabolism \cite{chen:04}.

\vspace{3mm}
{\noindent  \large \bf Discussion}
\vspace{1mm}

\noindent If flows across a membrane are governed by the ordinary diffusion
equation, they will reverse exactly, if the concentration gradient
across the membrane is reversed, regardless of the geometry of the
pores. Implicit in the use of the linear diffusion equation is the
point-particle assumption. Particles do not interact, and solutions
to the equations of motion are the same up to a factor for a dense
gas or a dilute gas, both in the bulk and in the pore.

If the diffusing particles have a size comparable to the pores
through which they are moving, the linear diffusion equation is no
longer applicable.  In the dilute limit, flows will still be
symmetric under reversal of the concentration gradient, for any pore
shape. But when concentrations are raised and
particles interact with each other and the walls of an asymmetric
pore, the symmetry is broken, and the diffusion rates from one side
or the other can differ.

Flux ratio theorems indicating 
a symmetry in diffusion across membranes have been proven for 
a variety of systems, e.g. with linear intra-membrane trapping kinetics 
for the diffusing particles \cite{stenn:81,bass:86} and even for 
multicomponent systems such as the one studied here \cite{mcnabb:90}.  
However, these theorems fail here because the trapping kinetics do not include the 
kinetic constraints felt by the spatially extended particles interacting with pores.

At first examination asymmetric diffusion might seem counterintuitive,
at odds with the Second Law.  But there is  no contradiction, as this is
a nonequilibrium effect, being driven here by increased
concentration on one side.  At equilibrium, detailed balance will hold,
and we expect fluctuations around equilibrium to be symmetric.
But a clogged membrane is held away from equilibrium, in a long-lived
metastable state.  A concentration gradient across the pore is required
to keep blocking particles in place, but this gradient can be made arbitrarily
small by lengthening the pore.

Asymmetric diffusion could be relevant to biological situations; rectification can
be generated in isothermal conditions, driven by only a small concentration
gradient. Asymmetry has long been
noted in movements of ions through channels in membranes, where the
asymmetry is often described as a form of rectification
\cite{hille:01}. Although ion channels can be complicated devices,
with voltage sensors and internal conformational changes, a
particularly simple pore design is the inward-rectifier potassium
channel; for a review see Lu \cite{lu:04}.  Early modeling of this
channel described the outward flow of K$^+$ ions through
the channel as being blocked by larger ions, which are present on
the inside of the cell and cannot pass completely through
\cite{armstrong:65,hille:78,vandenberg:87,matsuda:87}.  Recent
detailed imaging \cite{doyle:98}  of a potassium channel shows an
asymmetic pore, narrow towards the outside and wider towards the
inside, roughly like our model pores represented in
Figs.~\ref{fig:apparatus}{\it B} and \ref{fig:simulation_image}.

The early rectification models of Hille and Schwartz \cite{hille:78}
explicitly included the requirement that a blocker back out of the
channel to re-enable flow. However, these and other models emphasize
electrical effects, in particular intra-pore binding sites, to keep
the blockers in place in the pore.  Electrostatic interactions among
ions and possible binding sites will of course be very important,
but we have demonstrated that an asymmetric geometry alone can also
lead to rectification.  We suggest that the clogging effect
described in this paper could play a role in maintaining blocking
and rectification.  The inclusion of this effect may help clarify
discussions of ``multi-ion channels'' \cite{hille:78,lester:98}.
Geometry-induced asymmetric diffusion could occur in any physical
context that includes the interaction of shapes in restrictive
geometries, as exemplified by our experiment and its membrane model.
It is natural for large objects to get stuck and to block passage of
following objects; all that is required is a certain density and a
flow direction, no matter how small the flow. The simplicity of the
conditions generating the effect suggest that it might occur quite
commonly in biological systems, where membranes are ubiquitous and
molecules of all shapes and sizes are closely packed
\cite{goodsell:93}.
 Large changes in forward and backward diffusion
rates can be effected by small changes in pore geometry, or object
shape. The simple clogging phenomenon arising from asymmetric pores
suggests a possible evolutionary path from a purely mechanical
rectification that might occur in a pre-biotic vesicle, to the
sophisticated membrane ion channels present in living systems today.

\vspace{3mm}
{\noindent  \large \bf Materials and Methods}
\vspace{1mm}

{\bf \noindent Experiment.} A parallelepiped container is made of standard
microscope slides for the top and sides, and cut slides for the
ends. The bottom of the container is brass with the interior surface
machined with a pattern of 2 mm wide divots to deflect beads from
vertical motion, to help randomize the motion. The container is
divided into two chambers by a brass membrane whose thickness $D$
was varied (Fig.~\ref{fig:apparatus}{\it A}). Each membrane has 16
circular pores with a long large-diameter section (whose length
depends on the membrane thickness) and a concentric short 
small-diameter constriction (whose length is fixed, 1 mm), as shown in
Fig.~\ref{fig:apparatus}{\it B}.

Particles of two different sizes are obtained by repeated sieving of
soda lime glass beads (sphericity $>$ 95\%) with a broad size
distribution (Ceroglass GSR).  The ``large" beads (average mass 7.6
mg) have diameters 1.70-1.85 mm, small enough to fit into the 
large-diameter end of a membrane pore but too large to pass through the
narrow end of a pore. The ``small" beads (average mass 4.2 mg) have
diameters 1.40-1.55 mm, small enough so they can easily pass
completely through the pores, but large enough so that they have to
stay in single file in the large-diameter portion of the pore (see
Fig.~\ref{fig:apparatus}{\it B}).

In each run the chamber on one side of the membrane is loaded with
beads, and then the container is oscillated vertically at 30 Hz
with an acceleration amplitude 9.5$g$ ($g$ is the gravitational
acceleration). To the eye, the bead motion is vigorous and random.
Pictures such as those in Fig.~\ref{fig:apparatus}{\it C} and {\it
D} are taken at one-second intervals (1/2000 s exposure time) with
a digital imaging system (Kodak Motion Corder Analyzer SR 1000)
synchronized with the driving frequency.

The number of small beads passing through the membrane into an
initially empty chamber is counted as a function of time using an
image processing algorithm that is accurate to within one or two
beads for up to about 80 beads, as in Fig.~\ref{fig:apparatus}{\it
C}. For a larger number of beads, as in Fig.~\ref{fig:apparatus}{\it
D}, the algorithm undercounts the number of beads due to eclipsing
and grouping effects, but a correction factor obtained by weighing
the beads and checked by direct hand counting yields counts accurate
to 5\%.

\vspace{3mm}
{\noindent \bf Hard disk simulation.} The simulation uses a two-dimensional
container divided by a membrane into two chambers, each 3 units high
by 2 units wide, as shown in Fig.~\ref{fig:simulation_image}. The
membrane has nine tapered pores, 0.04 units wide on the large end
and 0.02 units wide on the small end. There are two species of elastic hard
disks, ``small" disks (diameter 0.018) and ``large" disks (diameter
0.022). As in the experiment, the small disks can pass through the
pores in either direction, while the large disks cannot pass through
the narrow end of a pore.  There are initially 1350 disks on the
right side, 20\% of which are the large disks; the left side is
initially empty.

All disks have the same mass and have initial unit velocities
uniformly distributed in random directions.  A period of at least 10
time units is allowed to pass before the pores are opened, to allow
the velocity distributions to thermalize; each simulation is
subsequently run for 500 time units. The work of Forster et al.~\cite{forster:04} 
shows that the dynamical entropy of such systems
is high, so that the Maxwellian distribution is rapidly achieved,
and preserved even in very restrictive geometries.

\vspace{3mm}
{\noindent \small 
We thank Josh Deutsch for a critical reading of the manuscript, and
Mark Bedau, Jim Bredt, Jim Crutchfield, Robert S. Eisenberg,
Doyne Farmer, Kristian Lindgren, and John McCaskill for helpful
comments. This research was partially funded by PACE (Programmable
Artificial Cell Evolution), and by ISCOM (Information Society as a
Complex System), European Projects in the EU FP5 and FP 6, IST-FET
Complex Systems Initiative.  RS and NP appreciate hospitality of the
Santa Fe Institute and the European Center for Living Technology.
The experiments were conducted at the University of Texas with the
support of the Sid Richardson Foundation.}


\begin{thebibliography}{10}

\bibitem{hille:01}
Hille, B. (2001) {\em Ion Channels of Excitable Membranes} (Sinauer Associates,
  Sunderland).

\bibitem{chen:93}
Chen, D.~P. \& Eisenberg, R.~S. (1993) {\em Biophys. J .} {\bf 65}, 727--746.

\bibitem{siwy:02}
Siwy, Z. \& Fuli\'nski, A. (2002) {\em Phys. Rev. Lett.} {\bf 89}, 198103.

\bibitem{siwy:05}
Siwy, Z., Kosi\'nska, I.~D., Fuli\'nski, A., \& Martin, C.~R. (2005) {\em Phys.
  Rev. Lett.} {\bf 94}, 048102.

\bibitem{siwy:06}
Siwy, Z.S., Powell, M.~R., Kalman, E., Astumian, R.~D., \& Eisenberg, R.~S.
  (2006) {\em Nano Lett.} {\bf 6}, 473 --477.

\bibitem{kosinska:05}
Kosi\'nska, I.~D. \& Fuli\'nski, A. (2005) {\em Phys. Rev. E} {\bf 72}, 011201.

\bibitem{kosztin:04}
Kosztin, I. \& Schulten, K. (2004) {\em Phys. Rev. Lett.} {\bf 93}, 238102.

\bibitem{cervera:05}
Cervera, J., Schiedt, B., \& Ram\'irez, P. (2005) {\em Europhys. Lett.} {\bf
  71}, 35--41.

\bibitem{cervera:06}
Cervera, J., Schiedt, B., Neumann, R., Maf\'e, S., \& Ram\'irez, P. (2006) {\em
  J. Chem. Phys.} {\bf 124}, 104706.

\bibitem{Bauer:06}
Bauer, W.~R. \& Nadler, W. (2006) {\em Proc. Natl. Acad. Sci.} {\bf 103},
  11446--11451.

\bibitem{smeets:06}
Smeets, R.~M.~M., Keyser, U.~F., Krapf, D., Wu, M.~Y., Dekker, N.~H., \&
  Dekker, C. (2006) {\em Nano Lett.} {\bf 6}, 89 --95.

\bibitem{kolomeisky:07}
Kolomeisky, A.B. (2007) {\em Phys. Rev. Lett.} {\bf 98}, 048105.

\bibitem{chou:99}
Chou, T. (1999) {\em J. Chem. Phys.} {\bf 110}, 606--615.

\bibitem{kim:05}
Kim, K.~S., Davis, I.~S., Macpherson, P.~A., Pedley, T.~J., \& Hill, A.~E
  (2005) {\em Proc. R. Soc. A} {\bf 461}, 273--296.


\bibitem{cleuren:06}
Cleuren, B., denBroeck, C.~Van, \& Kawai, R. (2006) {\em Phys. Rev. E} {\bf
  74}, 021117.

\bibitem{brilliantov:04}
Brilliantov, N.~V. \& P\"oschel, T. (2004) {\em Kinetic Theory of Granular
  Gases} (Oxford University Press, Oxford).

\bibitem{chen:04}
Chen, I. \& Szostak, J.W. (2004) {\em Proc. Natl. Acad. Sci.} {\bf 101}, 7965.

\bibitem{stenn:81}
Sten-Knudsen, O. \& Ussing, H.H. (1981) {\em J. Membrane Biol.} {\bf 63},
  233--242.

\bibitem{bass:86}
Bass, L., Bracken, A.J., \& Hilden, J. (1986) {\em J. Theor. Biol.} {\bf 118},
  327--238.

\bibitem{mcnabb:90}
McNabb, A. \& Bass, L. (1990) {\em IMA J. of Appl. Math.} {\bf 44}, 155--161.

\bibitem{lu:04}
Lu, Z. (2004) {\em Ann. Rev. Physiology} {\bf 66}, 103--129.

\bibitem{armstrong:65}
Armstrong, C.~M. \& Binstock, L. (1965) {\em J. Gen. Physiol.} {\bf 48},
  859--872.

\bibitem{hille:78}
Hille, B. \& Schwartz, W. (1978) {\em J. Gen. Physiol.} {\bf 72}, 409--442.

\bibitem{vandenberg:87}
Vandenberg, C.~A. (1987) {\em Proc. Natl. Acad. Sci.} {\bf 84}, 2560--2564.

\bibitem{matsuda:87}
Matsuda, H., Saigusa, A., \& Irisawa, H. (1987) {\em Nature} {\bf 325},
  156--159.

\bibitem{doyle:98}
Doyle, D.A., Cabral, J.~M., Pfuetzner, R.~A., Kuo, A., Gulbis, J.~M., Cohen,
  S.~L., Chait, B.~T., \& MacKinnon, R. (1998) {\em Science} {\bf 280}, 69--77.

\bibitem{lester:98}
Lester, H.~A. \& Dougherty, D.~A. (1998) {\em J. Gen. Physiol.} {\bf 111},
  181--183.

\bibitem{goodsell:93}
Goodsell, D.~S. (1993) {\em The Machinery of Life} (Springer-Verlag, New York).

\bibitem{forster:04}
Forster, Ch., Mukamel, D., \& Posch, H.A. (2004) {\em Phys. Rev. E} {\bf 69},
  066124.

\end{thebibliography}

\end{document}